\documentstyle[twocolumn,aps,amsmath,amssymb,graphicx,psfrag,prl]{revtex}
\def\eg{e.g.\hbox{}}
\def\etal{{\it et~al.\/}}
\def\ie{i.e.\hbox{}}

\def\ccbeta{\eta}
\def\ltsim{\lesssim}
\def\gtsim{\gtrsim}

\def\constant{\text{constant}}
\def\turn{{\text{turn}}}
\def\CSS{{\text{CSS}}}
\def\DSS{{\text{DSS}}}

\def\Bizon{Bizo\'{n}}

\begin{document}
\title{A New Transition between Discrete and Continuous Self-Similarity
       in Critical Gravitational Collapse}
 \author{Christiane Lechner$^{(1,2)}$,
       Jonathan Thornburg$^{(1,2)}$,
       Sascha Husa$^{(2)}$,
       and Peter C.~Aichelburg$^{(1)}$}
 \address{$^{(1)}$
          Institut f\"{u}r Theoretische Physik,
          Universit\"{a}t Wien,                                          \\
          Boltzmanngasse~5, A-1090 Wien, Austria                         \\
          $^{(2)}$
          Max-Planck-Institut f\"ur Gravitationsphysik,
          Albert-Einstein-Institut\\
          Am M\"uhlenberg 1, D-14476 Golm, Germany                     \\
          }
 
%
%
\date{Dec.~7 2001}
\maketitle


\begin{abstract}
We analyze a bifurcation phenomenon associated with critical
gravitational collapse in a family of self-gravitating SU(2) $\sigma$-models.
As the dimensionless coupling constant decreases,
the critical solution changes from discretely self-similar (DSS)
to continuously self-similar (CSS). 
Numerical results
provide evidence for a bifurcation which
is analogous to a heteroclinic loop bifurcation in dynamical systems,
where two fixed points (CSS) collide with a limit cycle (DSS) in phase
space as the coupling constant tends to a critical value.
\end{abstract}



Gravitational collapse at the threshold of black hole formation
ties together many of the fundamental issues of general relativity, such
as the global aspects of solutions, the structure of singularities arising
from regular initial data and the cosmic censorship hypothesis~
\cite{Penrose-strong-cosmic-censorship}.
By fine-tuning initial data for a gravitating massless scalar field
to the boundary between eventual dispersal and complete collapse,
Choptuik~\cite{Choptuik-1993-self-similarity} found phenomena reminiscent
of criticality associated with phase transitions in statistical physics,
such as universality and scaling (e.g. of the black hole mass).
Considerable qualitative understanding has been gained by explaining
critical collapse in terms of a single unstable mode of the universal
critical solution. This solution is understood as an intermediate attractor
located in a codimension-one stable hypersurface in phase space,
separating data which do or do not form black holes.
Critical collapse has by now been studied in a number of matter models --
in all of these the physics of the threshold of black hole formation
was found to be governed by symmetry: the critical solution exhibits
either continuous or discrete self-similarity, or staticity or
periodicity in time~\cite{Gundlach-1999-critical-phenomena-review}.

In this paper, we study numerically a simple model with a single scalar
field in spherical symmetry which exhibits CSS critical behavior at
small coupling constants, and DSS critical behavior at large ones.
At intermediate coupling constants we observe a
competition between CSS and DSS solutions giving rise to a new phenomenon:
within an approximately-DSS critical evolution we find several episodes of
approximate CSS. Our main focus here is on the interpretation of this
observation in terms of an analogy to a heteroclinic loop bifurcation
in finite dimensional dynamical systems, at which the limit cycle
(DSS) merges
with two fixed points (the CSS solution and its negative).
Apart from the bifurcation itself, the model also shows other interesting
features, such as the existence of a stable (with respect to linear spherical
perturbations) self-similar solution for some finite range of the coupling,
and a ``suppression'' effect for the CSS solution in critical searches
which we interpret as ``shielding'' by an apparent horizon.

Our results are based on the direct numerical construction of the 
CSS and DSS solutions, a linear perturbation analysis of the CSS
solutions, and comparison with critical evolutions. The latter are
defined by considering a 1-parameter family of initial data
$\phi = \phi_p(u_0,r)$, such that (say) for small values of~$p$
the field eventually disperses (as determined by a numerical
evolution~\cite{Husa-etal-2000-sigma-model-DSS-criticality}), while
for large values of $p$ it eventually forms a black hole (diagnosed by
the appearance of an apparent horizon).  We use a binary search in~$p$
to numerically approximate the critical solution at the
threshold of black hole formation.
We refer to such fine-tuned numerical solutions as near-critical
evolutions, and our results are taken from initial data
which are fine-tuned to the same tolerance $\delta p/p < 10^{-14}$.


The self-gravitating SU(2) $\sigma$-model \cite{Misner-1978-harmonic-maps}
under investigation
is a wave map from spacetime to the target manifold $S^3$ with the
standard metric. The so called Hedgehog ansatz of spherical symmetry
leaves a single matter field $\phi(u,r)$ coupled to gravity:  
\begin{equation}
  \square\phi
      =\frac{\sin (2\phi )}{r^{2}}
                                                                \,\text{,}
                                                                \label{eqn-phi} 
\end{equation}
where $\square$ is the spacetime wave operator.  Our geometric setup,
numerical evolution scheme, and convergence tests are described in a previous
paper~\cite{Husa-etal-2000-sigma-model-DSS-criticality}.
In particular, we use  retarded Bondi-like coordinates $(u,r)$ with
metric functions $\beta (u,r)$ and $V(u,r)$. 
Suitable combinations of Einstein's equations lead to
\begin{equation}
\beta'  = \frac{\ccbeta }{2}r(\phi')^2
\text{,}
\quad
V' = e^{2\beta}(1-2\ccbeta \sin^2 \phi)
                                                                \,
\text{,}
                                                        \label{eqn-beta'-V'}
\end{equation}
where prime denotes the derivatives with respect to $r$,
and $\eta$ the dimensionless coupling constant.
The hypersurface equations~\eqref{eqn-beta'-V'}
and the matter field equation~\eqref{eqn-phi} suffice to evolve
all the dynamical fields $V$, $\beta$, and $\phi$.

\begin{figure}[htb]
\begin{center}
\begin{psfrags}
\psfrag{Delta}[c][c][1][-90]{$\Delta$}
\psfrag{eta}[c][c]{$\eta$}
\includegraphics[width=40ex]{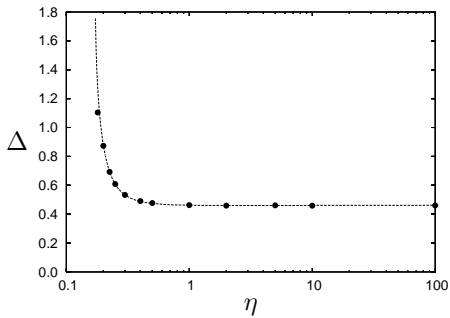}
\end{psfrags}
\end{center}
\caption[DSS echoing period $\Delta$]
        {
        The DSS echoing period $\Delta$ 
        is shown as a function of the coupling constant $\eta$,
        computed from dynamical evolutions (dots) and from
        direct construction (line).
        Our dynamical evolution is only approximately DSS
        at its lowest $\eta$ value ($\eta = 0.18$), so we
        can only determine $\Delta$ approximately there.
        }
\label{fig:Delta(eta)}
\end{figure}

For vanishing coupling  ($\eta = 0$), the theory describes a $\sigma$-model
on a fixed background. Taking this background as Minkowski space,
\Bizon{} \etal{}~\cite{Bizon-Chmaj-Tabor-1999-sigma-3+1-evolution} and
Liebling \etal{}~\cite{Liebling-Hirschmann-Isenberg-1999-sigma-critical}
find a CSS critical solution at the threshold of singularity formation.
\Bizon{}~\cite{Bizon-1999-existence-of-self-similar-sigma-CSS-solutions} and
\Bizon{} and Wasserman~\cite{Bizon-Wasserman-2000-CSS-exists-for-nonzero-beta}
have shown that for each $0 \leq \eta < 0.5$, a countably infinite family
of CSS solutions exists, indexed by the number of nodes in
$\phi(u=\constant,r) - \pi/2$.
In the limit $\eta \rightarrow \infty$
Liebling~\cite{Liebling-inside-global-monopoles} finds DSS critical
collapse at the threshold of black-hole formation.
In Ref.~\cite{Husa-etal-2000-sigma-model-DSS-criticality}
we find that for $\eta \gtsim 0.2$ the system shows ``exact'' DSS
critical collapse, but for $0.18 \le \eta \ltsim 0.2$ we see
only approximate DSS behavior; furthermore the period~$\Delta$ exhibits
a sharp rise as the coupling decreases from~$0.5$ to~$0.18$
(Fig.~\ref{fig:Delta(eta)}).  These results suggest a transition
from CSS to DSS critical collapse somewhere in the range
$0 < \eta \ltsim 0.18$, which we identify and discuss in the present
paper.

For $0 \leq \eta < 0.5$ we have numerically constructed the CSS
solutions discussed in
\cite{Bizon-Wasserman-2000-CSS-exists-for-nonzero-beta} as an
ODE-eigenvalue problem.
We have also performed a linear stability analysis of the CSS solutions;
we find that the first excitation (one node),
which we refer to below as ``the'' CSS solution, has precisely
one unstable mode.  This CSS solution takes the form
$\phi = \pm \phi_\CSS(z;u^*_\CSS)$, where $z = r/(u^*_\CSS-u)$, with a single
free parameter $u^*_\CSS$ giving the retarded time of the accumulation
point. The sign ambiguity is a consequence of Einstein's equations and
the field equation~\eqref{eqn-phi} being invariant under $\phi \to -\phi$.

For $\eta \geq 0.1726$ we have also explicitly constructed the
``Choptuon'' DSS solution via a pseudospectral method following
the lines of Gundlach~\cite{Gundlach-1996-understanding-critical-collapse}
(see \cite{Lechner-DSS-forthcoming}).
The DSS solution takes the form
$\phi = \phi_\DSS(\tau,z;u^*) = \phi_\DSS(\tau+n\Delta,z;u^*)$,
where $n$ is any integer, $\Delta$ is the DSS period,
$z$ is again given by $r/(u^*-u)$, and $\tau = - \ln(u^*-u)$.  
As $\eta$ decreases $\Delta$
rises sharply (Fig.~\ref{fig:Delta(eta)}).
Also, a rapidly increasing number of Fourier components is required
to accurately represent the Choptuon, and the construction algorithm
becomes increasingly ill-conditioned.  
Below we will give further arguments suggesting that the
DSS Choptuon ceases to exist somewhat below the lower limit of our
numerical construction.



For very small couplings $\eta \ltsim 0.1$ the stable CSS ground state
causes a generic class of initial data to collapse to naked singularities.
Here we focus on the transition from CSS to DSS in critical collapse
at the threshold of {\em black hole\/} formation:  We therefore restrict
our attention to $\eta \geq 0.1$, where we find only dispersal and
black hole (apparent horizon) formation as generic end-states.

For $0.1 \leq \eta \ltsim 0.14$ we find CSS critical collapse, while
for large couplings $\eta \gtsim 0.2$ we have previously found DSS
critical collapse~\cite{Husa-etal-2000-sigma-model-DSS-criticality}.
In both ranges we observe scaling of the black hole mass for
supercritical initial data and of the maximum central Ricci scalar
for subcritical initial data. In the CSS regime $0.1 \leq \eta \ltsim 0.14$
and in the DSS regime for $\eta \gtrsim 0.2$ the 
critical exponents are approximately constant (within a few percent):
$\gamma_\CSS \approx 0.18$ and $\gamma_\DSS \approx 0.11$.


In the transition regime $0.14 \ltsim \eta \ltsim 0.2$, we find
that critical solutions show a new phenomenon which we call
``episodic self-similarity'': The field configuration closely
approximates CSS behavior on large parts of the slice for a finite time,
then departs and returns to CSS again.
This cycle repeats several times
before the evolution either leads to black hole formation or dispersal.
We find that $\phi \approx +\phi_\CSS$ and $\phi \approx -\phi_\CSS$
episodes always alternate.  The accumulation times $u^*_\CSS$
increase from one CSS episode to the next.


\begin{figure}[htb]
\begin{center}
\begin{psfrags}
\psfrag{-log(u* - u)}[c][c]{$- \ln (u^* - u)$}
\psfrag{RMS distance in phi}[c][c]{RMS distance in $\phi$}
\psfrag{max 2m/r}[c][c]{max $2m/r$}
\includegraphics[width=87mm]{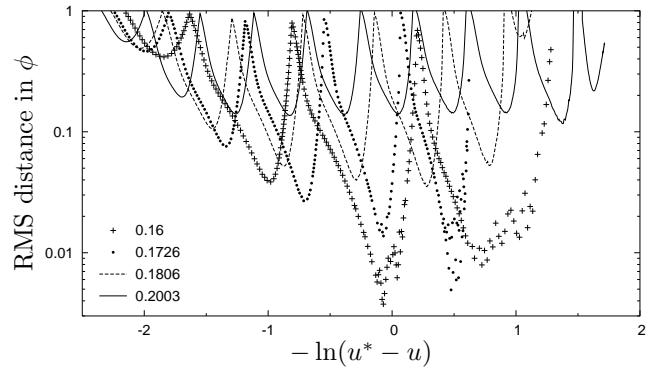}
\end{psfrags}
\end{center}
\caption[Figure showing CSS-DICE fits at various coupling constants]
        {
        This figure shows the distance ($r^2$-weighted RMS difference
        in $\phi$, taken between the origin and the self-similarity
        horizon, within each near-critical-evolution slice) 
        between the CSS solution and the critical solution, for
        $\eta = 0.16$, $0.1726$, $0.1806$, and $0.2003$.
        (The choice of $u^*$, and thus the horizontal coordinate,
        is somewhat arbitrary for $\eta = 0.16$.)
        }
\label{fig:css-dice}
\end{figure}

In order to study episodic self-similarity quantitatively, we have fitted
numerical near-critical evolutions against our explicitly-constructed CSS solutions
(fitting the CSS parameter $u^*_\CSS$ independently at each 
near-critical-evolution
slice).  Figure~\ref{fig:css-dice} shows these fits for a range of
coupling constants.  The repeated close approaches of the near-critical
evolutions to the CSS solutions are clearly visible; the approaches
become closer and closer and the time spent in the neighborhood
(\ie{}~within a given distance) of the CSS solution increases
as $\eta$ is decreased.

In the range where episodic CSS occurs we also observe approximate DSS
behavior. It is therefore interesting to compare the near-critical evolution
to the explicitly constructed DSS solution.

\begin{figure}[]
\begin{center}
\begin{psfrags}
\psfrag{DSS tau (= Delta * phase in cycles)}[c][c]{DSS $\tau$}
\psfrag{RMS distance in phi}[c][c]{RMS distance in $\phi$}
\psfrag{max 2m/r}[c][c]{max $2m/r$}
\includegraphics[width=87mm]{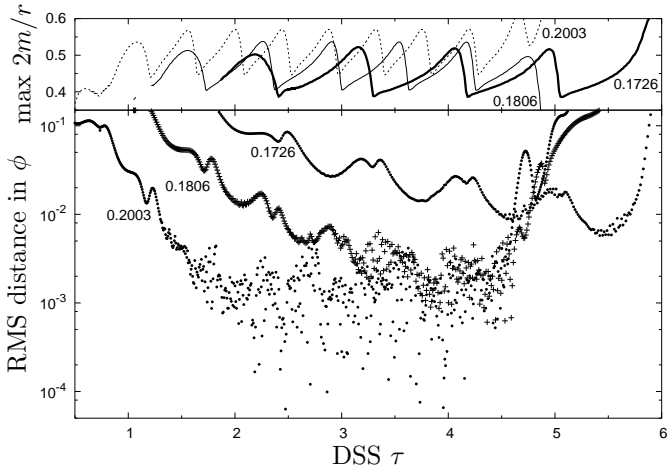}
\end{psfrags}
\end{center}
\caption[Figure showing DSS-DICE fits at various coupling constants]
        {
        Best fits between the DSS Choptuon
        and the critical solution are shown for $\eta = 0.1726$, $0.1806$,
        and $0.2003$.  The lower subplot shows the
        distance (same definition as in figure~\protect\ref{fig:css-dice})
        between the DSS Choptuon and the corresponding best-fitting
        slices of the near-critical evolution, as a function of 
        $\tau$.  The upper subplot shows the maximum of
        $2m/r$ within the same slices of the critical evolution.
        $\tau$ is only defined up to an arbitrary integer multiple
        of $\Delta$ at each coupling constant.
        }
\label{fig:dss-dice}
\end{figure}


Figure~\ref{fig:dss-dice} shows fits of numerical near-critical evolutions
against the explicitly-constructed DSS solutions (finding best-fitting
pairs of slices between the near-critical evolutions and the DSS solutions)
for several coupling constants where DSS exists. Notice that as $\eta$
decreases, the near-critical solution's approach to the DSS solution
becomes slower, and the closest approach becomes less close.
The time intervals in $\tau$ from approaching the Choptuon within an RMS error
of $\sim 0.1$ (which is where the curves in Fig.~\ref{fig:dss-dice} starts)
to the start of the departures are however roughly equal.
This and the slow approach account for the (only) approximate DSS behavior 
of near-critical evolutions already observed in 
\cite{Husa-etal-2000-sigma-model-DSS-criticality} for $0.18 \le \eta
\ltsim 0.2$.

\begin{figure}[]
\begin{center}
\begin{psfrags}
\psfrag{DSS phase (cycles)}[c][c]{DSS phase (cycles)}
\psfrag{RMS distance in phi}[c][c]{RMS distance in $\phi$}
\includegraphics[width=87mm]{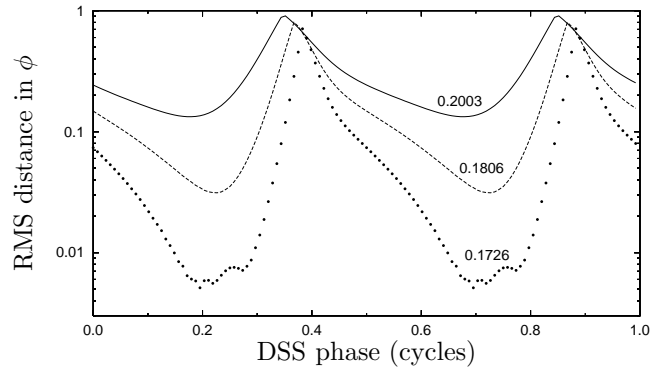}
\end{psfrags}
\end{center}
\caption[Figure showing CSS-DSS fits at various coupling constants]
        {
        The distance (same definition as in figure~\protect\ref{fig:css-dice})
        between the CSS and DSS solutions is shown as a function of DSS
        phase (measured in orbits around the DSS Choptuon), for
        $\eta = 0.1726$, $0.1806$, and $0.2003$.  The origin of
        the DSS phase scale is arbitrary at each coupling constant.
        }
\label{fig:css-dss}
\end{figure}

Comparing Figs. \ref{fig:css-dice} and \ref{fig:dss-dice} one infers that
as the critical evolution is attracted to the DSS solution it comes
periodically close to the CSS solution, which implies that the DSS and CSS
solutions must themselves be close.
Figure~\ref{fig:css-dss} shows fits between the CSS solutions and
the explicitly-constructed DSS solutions (again fitting the CSS 
parameter $u^*_\CSS$ independently at each slice).  There are two close 
approaches
within each DSS cycle, corresponding to the two sign choices
$\phi = \pm \phi_\CSS$. Note that the close approaches become closer as
$\eta$ decreases.


Combining our results, we conjecture the following bifurcation
scenario in the language of dynamical systems: for large couplings
$0.2 \ltsim \eta < 0.5$, the limit cycle representing $\phi_\DSS$
in phase space and the two fixed points $\pm \phi_\CSS$ lie far apart
(in some suitable norm).  As $\eta$ is decreased, the limit cycle and
the CSS points move closer and finally merge at some $\eta_c \approx 0.17$.
In this limit the DSS solution becomes a heteroclinic orbit connecting
the CSS fixed points and the period of the limit cycle tends to infinity
(see below).
For $\eta < \eta_c$, DSS ceases to exist.

We conjecture that the DSS solution still plays the role 
of a critical intermediate attractor even at coupling constants
just slightly larger than $\eta_c$, where the CSS solutions lie very close to
the DSS cycle.
An evolution, which is tuned to evolve towards the DSS-CSS region is carried
along by the flow of the DSS cycle to periodically come close to the 
CSS fixed points. We have numerical evidence that the periodical turning away
from CSS is dominated by the unstable mode of CSS.

When the DSS evolution
is close to one of the CSS fixed points in phase space we can expand
the field in terms of linear perturbations around the CSS solution.
The departure from CSS must thus happen via the unstable mode of CSS.
The amplitude for this mode grows from an initial amplitude $A_0$
to some fixed amplitude (still in the linear regime) in a
time $T = - (1/\lambda) \ln (A_0) + \constant$, where $\lambda$ denotes the
eigenvalue of the unstable mode of the CSS solution.  
Since this happens twice
in a DSS cycle, we can write the total duration of the DSS cycle as
$\Delta = 2T + T_{\turn}$, where $T_\turn$ denotes the time spent in
the (nonlinear) turnover from one of $\pm \phi_\CSS$ to the other.
From our linear perturbation analysis of the CSS solution we find that
$\lambda \approx 5.14$ is only slowly varying for $\eta$ near $\eta_c$
\cite{Lechner-PhD}.  If we assume 
that $A_0 \sim \eta - \eta_c$ for $\eta$ near $\eta_c$
and that 
$T_\turn$ is roughly constant, we have
$\Delta = - (2/\lambda) \, \ln (\eta - \eta_c) + \textrm{const}$.
To test this prediction, we have fitted the $\Delta(\eta)$ values
shown in Fig.~\ref{fig:Delta(eta)} to the 3-parameter functional
form $f(\eta) = -a \ln(\eta - \eta_c) + b$ in the range 
$\eta \in [0.1726,0.195]$. 
The fit is very good, with a maximum relative error of $0.3\%$.
We obtain 
$\eta_c \simeq 0.17$, which is consistent with what we expect from the 
raise in the number of relevant Fourier coefficients, 
and $a \sim 2/\lambda$ with a relative error of $\sim 7 \%$. Given
the fact that we neglected higher order terms and the variation of $\lambda$,
the fitted value for $a$ is remarkably close to the theoretically predicted
one.

\begin{figure}[htb]
\begin{center}
\begin{psfrags}
\psfrag{eta}[c][c]{$\eta$}
\includegraphics[width=87mm]{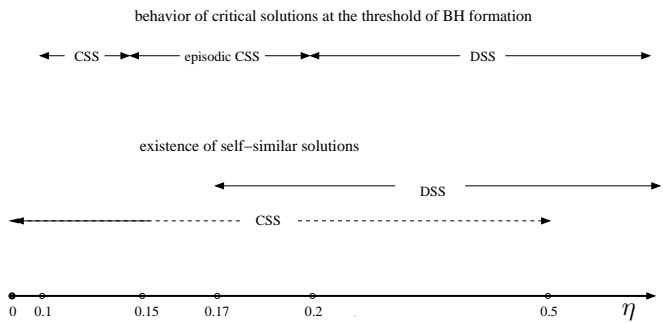}
\end{psfrags}
\end{center}
\caption[Overview of phenomenology at various coupling constants]
        {
        This figure shows the different phenomenology we observe
        at various couplings $\eta$. The dashed line denotes
        CSS solutions, that contain marginally trapped surfaces.
        }
\label{fig:behavior-overview}
\end{figure}

Figure~\ref{fig:behavior-overview} gives a schematic overview of all our
observations.  




We also observe episodic CSS  at couplings  $0.14 \lesssim \eta < \eta_c$,
where we believe the DSS solution ceases to exist.
What is tuned out in such  a critical search? In the language of dynamical
systems the answer could be that below the critical coupling the cycle
of DSS is broken \ie{} it does not close,  but the flow still defines locally
an invariant manifold of codimension one. It is this manifold which
separates black hole formation from dispersal.  
The mass scaling from supercritical searches in the intermediate
regime $0.14 \ltsim \eta < 0.2$ has not been conclusive so far.
Observed deviations from a simple scaling law 
require further investigation.

Another important question is: Why does the CSS solution cease to 
be a critical solution for black hole formation for larger couplings?
Our stability analysis shows that CSS has a single
unstable mode up to $\eta = 0.5$, which in principle could be tuned out in
a critical search. 
We believe that the answer to this is related to the observation by
\Bizon{} and Wasserman
\cite{Bizon-Wasserman-2000-CSS-exists-for-nonzero-beta}, that this
solution contains a spacelike hypersurface of marginally trapped surfaces
outside the backwards light cone of the culmination point for $\eta > 0.152$.
We find that numerical evolutions for $\eta = 0.2$ 
with initial data that are close to the CSS solution
inside the backwards lightcone and are asymptotically flat outside, very
quickly develop an apparent horizon and thus become a black hole. If this is
the generic behavior, then the CSS solution can not lie on the boundary of
black hole formation.  


Summing up, the SU(2) $\sigma$-model shows CSS critical
behavior for small and DSS for large values of the coupling constant.
In the transition region
we observe
episodic CSS behavior. We have strong evidence that the CSS/DSS
transition is the infinite dimensional analog of a global heteroclinic
bifurcation, which is quite different
from previously reported bifurcations in self-similar critical collapse, which
were found to be characterized by a change of stability (see \eg{} 
\cite{Liebling-Choptuik-CSS-DSS}).
In particular, analogies to finite dimensional dynamical systems
pictures have proven essential in interpreting critical collapse
(see e.g. Ref.~\cite{Gundlach-1999-critical-phenomena-review}
for an overview), and we believe that the bifurcation picture discussed here
will  stimulate further insights into critical gravitational collapse.
It would be interesting to see whether episodic CSS occurs in the
critical collapse of different matter models, or also in completely 
different physical systems.
Details of our methods and results, some of which could only be mentioned
here 
will be published in a forthcoming paper.

This work has been supported by
the Austrian 
FWF
(project P12754-PHY), the Fundacion Federico,
the Alexander von Humboldt Foundation,
and G.~Rodgers and J.~Thorn [J.T.].
We thank Michael P\"urrer for his contributions to our numerical
evolution code~\cite{Husa-etal-2000-sigma-model-DSS-criticality},
Piotr \Bizon{} for many useful discussions and for sharing research
results in advance of publication,
and Carsten Gundlach and Jos\'e M.~Mart\'\i{}n-Garc\'\i{}a
for stimulating discussions
and for drawing our attention to the analysis leading to the logarithmic
divergence of $\Delta$.


\bibliography{jt}

\end{document}